\documentclass[runningheads]{llncs}
\usepackage[T1]{fontenc}
\usepackage{graphicx}
\usepackage[hidelinks]{hyperref}
\usepackage{color}

\usepackage{calc,tikz,tikz-3dplot}
\usetikzlibrary{arrows.meta}
\usetikzlibrary{backgrounds}
\usetikzlibrary{decorations.markings}
\usetikzlibrary{positioning}
\usetikzlibrary{shapes.geometric}
\usepackage{xcolor}
\definecolor{softorange}{HTML}{ff7f0e}
\definecolor{softgreen}{HTML}{2ca02c}
\definecolor{softblue}{HTML}{1f77b4}
\definecolor{softred}{HTML}{d62728}
\definecolor{red}{RGB}{220, 0, 0}
\definecolor{white}{RGB}{255, 255, 255}
\definecolor{darkgreen}{RGB}{0, 130, 0}
\definecolor{lightcyan}{RGB}{127, 255, 255}
\definecolor{purecyan}{HTML}{00ffff}
\definecolor{purepink}{HTML}{ff00ff}
\definecolor{matlabblue}{rgb}{0, 0.4470, 0.7410}
\definecolor{bordeaux}{HTML}{6E001D}
\hypersetup{
    colorlinks = true,
    citecolor = bordeaux,
    linkcolor = bordeaux
}
\usepackage{textcomp}
\usepackage{amsmath}
\usepackage{amssymb}
\renewcommand{\vec}[1]{\boldsymbol{#1}}

\DeclareMathOperator{\dir}{dir}
\DeclareMathOperator{\valid}{valid}
\usepackage{tabularx}
\usepackage{stackrel}
\usepackage{marvosym} 
\begin{document}
\title{A Spatiotemporal Model for Precise and Efficient Fully-automatic 3D Motion Correction in OCT}
\titlerunning{A Spatiotemporal Model for 3D Motion Correction in OCT}
%
\author{Stefan Ploner\inst{1,2\textrm{(\Letter)}} \and
Siyu Chen\inst{2} \and
Jungeun Won\inst{2} \and
Lennart Husvogt\inst{1} \and
Katharina Breininger\inst{1} \and
Julia Schottenhamml\inst{1} \and
James Fujimoto\inst{2} \and
Andreas Maier\inst{1}
}
%
\authorrunning{S. Ploner et al.}
\institute{Friedrich-Alexander-Universität Erlangen-Nürnberg, Erlangen, Germany
\and
Massachusetts Institute of Technology, Cambridge, MA, USA\\
\email{stefan.ploner@fau.de}
}
\maketitle              
\begin{abstract} 
Optical coherence tomography (OCT) is a micrometer-scale, volumetric imaging modality that has become a clinical standard in ophthalmology.
OCT instruments image by raster-scanning a focused light spot across the retina, acquiring sequential cross-sectional images to generate volumetric data.
Patient eye motion during the acquisition poses unique challenges:
Non-rigid, discontinuous distortions can occur, leading to gaps in data and distorted topographic measurements.
We present a new distortion model and a corresponding fully-automatic, reference-free optimization strategy for computational motion correction in orthogonally raster-scanned, retinal OCT volumes.
Using a novel, domain-specific spatiotemporal parametrization of forward-warping displacements, eye motion can be corrected continuously for the first time.
Parameter estimation with temporal regularization improves robustness and accuracy over previous spatial approaches.
We correct each A-scan individually in 3D in a single mapping, including repeated acquisitions used in OCT angiography protocols.
Specialized 3D forward image warping reduces median runtime to < 9 s, fast enough for clinical use.
We present a quantitative evaluation on 18 subjects with ocular pathology and demonstrate accurate correction during microsaccades.
Transverse correction is limited only by ocular tremor, whereas submicron repeatability is achieved axially (0.51 \textmu m median of medians), representing a dramatic improvement over previous work.
This allows assessing longitudinal changes in focal retinal pathologies as a marker of disease progression or treatment response, and promises to enable multiple new capabilities such as supersampled/super-resolution volume reconstruction and analysis of pathological eye motion occuring in neurological diseases.

\keywords{Optical coherence tomography \and Motion compensation \and Non-rigid registration \and Forward warping.}
\end{abstract}

\section{Introduction} \label{intro}

\begin{figure}[t]%
    \centering%
    \includegraphics{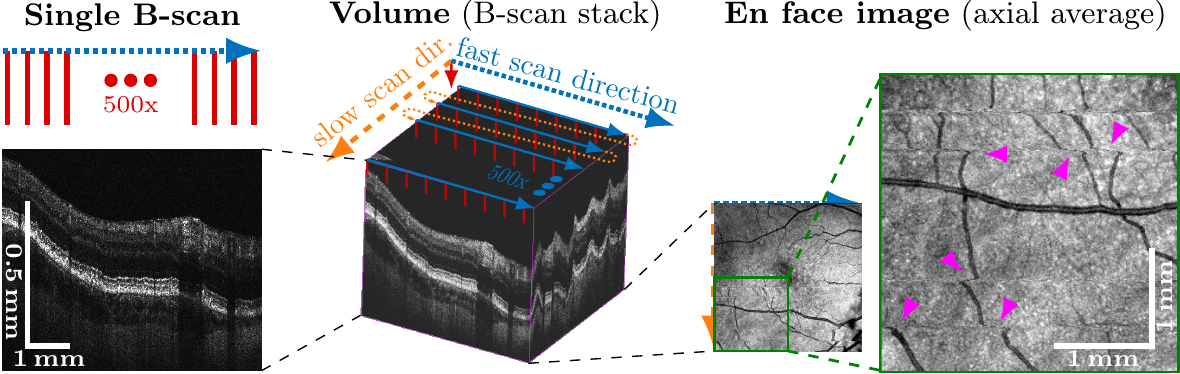}%
    \caption{A volume is composed of A-scans (red lines, in \emph{axial} direction). The \emph{transverse} scanned trajectory indicated in the volumetric scan, consisting of B-scans (blue arrows) and flyback (orange lines), defines continuity of image distortion. Microsaccade discontinuities (pink markers) are best visible after axial averaging, in \emph{en face} images.}%
	\label{fig:oct-scanning}%
\end{figure}%
\begin{figure}[t]%
	\centering%
    \includegraphics{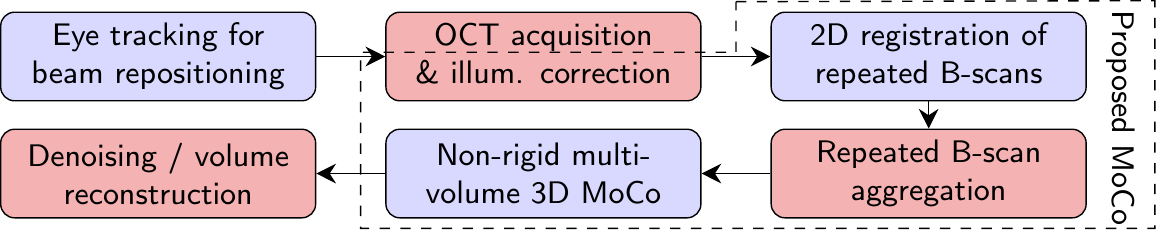}%
	\caption{Classic OCT processing pipeline. Steps related to motion correction (MoCo) in light blue, steps related to signal reconstruction and illumination (illum.) correction in red. Repeated B-scan registration is limited to 2D and increases gap size (see \figurename~\ref{fig:rep-bscans}B).}%
	\label{fig:oct-processing}%
\end{figure}
Imaging the eye is particularly challenging because three types of involuntary eye movements continuously occur even during fixation~\cite{MartinezConde2004}:
\emph{Microsaccades} (occasional fast movements lasting < $\sim$25~ms),
\emph{drift} (slow, random walk-like motion causing distortion throughout), and \emph{tremor} (aperiodic transverse vibration below the resolution of non-adaptive optics retinal imaging).
Fixation capabilities decrease with age and in pathology.
Optical coherence tomography (OCT) is a non-invasive 3D imaging modality and standard of care in ophthalmology~\cite{Huang1991}. By raster-scanning a laser beam across the retina, OCT assembles a volume of quasi-instantaneous depth profiles of backscattering (A-scans). A line of A-scans forms a 2D cross-sectional image (B-scan) with minimal distortion due to a millisecond acquisition (\figurename~\ref{fig:oct-scanning}).
In contrast, volume acquisition requires seconds and is correspondingly more distorted in slow scan direction.
Microsaccades appear as discontinuities and cause gaps if performed opposite to the slow scan direction.
OCT angiography (OCTA) visualizes microvasculature by performing repeated B-scans at the same retinal location and detecting motion contrast from moving blood cells~\cite{Spaide2018}. However, scan time and
distortion are further increased.

Previous motion correction methods are based on a three-step procedure outlined in \figurename~\ref{fig:oct-processing}.
Eye tracking compensates motion and, thus, gap size during acquisition, but is insufficient as typically only transverse motion is corrected, and accuracy is fundamentally limited by latency
in the range of a microsaccade duration~\cite{Schwarzhans2020}.
Secondly, repeated B-scans are affinely registered.
This step is limited to correcting in-plane (2D) motion by its 2D nature~\cite{Spaide2018}.
Subsequently, repeated B-scans are aggregated by averaging (OCT) or variance computation (OCTA).
Lastly, two or more volumes are coregistered, allowing latency-free 3D distortion correction.
Additional scans increase acquisition time, but enable data fusion for gap filling and signal-to-noise ratio enhancement~\cite{Spaide2018}.

For the volumetric correction,
typically a pair of volumes with orthogonally oriented B-scans is used. This allows correcting distortion along the slow scan direction with the B-scans of the orthogonal volume and vice-versa. The main challenges are lack of a distortion-free reference, and the spatial discontinuity in distortion due to raster-scanning.
Brea et al.\ reviewed existing methods and concluded that
current models are insufficient to accurately correct fine features
\cite{Brea2019}.
More recently, Ploner et al.\ integrated utilization of OCTA data~\cite{Ploner2021} into the method of Kraus et al.~\cite{Kraus2014},
which is in commercial use.
A special feature of these methods is that they are reference-free: displacement fields for each scan are jointly estimated in an axial, followed by a 3D, iterative optimization. This avoids propagation of drift motion from a reference to the result.
Axial undistortion is achieved reliably, but correction is limited by use of backward warping:
B-scan shearing must be estimated and compensated in the initial axial optimization, but this introduces inaccuracies because it lacks correction of transverse motion.
Discontinuities in the distorted inputs (pink markers in \figurename~\ref{fig:oct-scanning}) may cause spurious gradients in the similarity metric (during optimization).
Furthermore, backward-warped displacements are defined at \emph{target} grid voxels, limiting continuity regularization to target grid axes.
However, B-scans acquired during slow scan direction motion or torsional eye motion (around the optical axis) are no longer axis-aligned with the target grid.
This must be compensated by discontinuities in the displacement fields which \emph{violate} the regularizer.
Consequently, issues occur in datasets with head tilt (many volumes or widefield), and misregistration persists especially near slow scan direction microsaccades.
While, e.\,g., the method of Athwal et al.\ compensates torsional motion~\cite{Athwal2021}, its local undistortion is limited (see discussion).
The more promising forward warping was avoided for iterative OCT motion correction because the necessary scattered data interpolation was assumed computationally infeasible for clinical settings.

\textbf{Contribution.}
We present the first spatiotemporal motion model and a corresponding fully-automatic, reference-free optimization strategy for volumetric distortion correction in OCT.
Our approach introduces numerous \textit{advances} over the state of the art:
By using a displacement field \textit{parametrization with respect to time}, discontinuities in optimization are remedied. Compared to prior work, a \textit{drastically reduced parameter density} suffices without compromising accuracy, and \textit{smoothness regularization} no longer contradicts with discontinuities.
Two features are critical to enable time-continuous modeling:
First, in order to infer consistent displacements despite distinct A-scan locations, motion is described for the eye as a whole, by a spatially rigid 3D transform. Remaining change is assumed to be of temporal origin.
Secondly, to be able to derive the displacement of each A-scan based on its acquisition time, the transform must be defined at the input voxels, i.\,e.\ formulated via \textit{forward warping}.
While forward warping~(FW) is fundamentally more complex to compute than backward warping, it enables image warping using a \textit{single, direct mapping}. This allows estimation of \textit{all parameters} in a single optimization, towards the (iteratively) \textit{fully corrected} orthogonal dataset.
A further advantage of FW is its \textit{correct handling of overlaps}, which enables individual registration of repeated B-scans during volume registration. This not only removes the necessity for prior registration and intermediate averaging of B-scan repeats, but more importantly, enables their \textit{correction in all three spatial dimensions}. Besides improving accuracy, this can substantially reduce gap size compared to prior in-plane registration methods.
Finally, FW enables direct integration of \textit{head tilt compensation}, effective \textit{compensation of displacement bias} from resampling effects, and gradient computation in the iteratively \textit{motion-corrected targets}, thereby bypassing spurious gradients in the input volumes.
Our model is suited for volume-pair, many-volume and widefield imaging, and is more reliable, dramatically more accurate, and converges faster.
As discussed at the end, a range of new applications is enabled.
Lastly, by exploiting the structure of possible distortion in the model, we formulate a domain-specific, \textit{separable forward interpolation scheme}
to enable
iterative optimization in clinically feasible runtimes. While adding all described advantages,
previously reported runtimes are reduced by a factor of 5.

\section{Methods}

Our approach jointly motion corrects two or more volumes with orthogonal B-scans.
The log-scale B-scans are preprocessed with a radius 1\,px median filter
and factor 2 axial downsampling.
OCT volumes are displayed in a grid where x,\,y displacements are within the coronal plane and z corresponds to depth.
We assume the image to be distorted by the 3D affine, temporally varying transform
\begin{equation} \label{eq:transform}
	\mathcal T(\vec x, \tau, \vec p) = \begin{pmatrix}
		\cos(-\alpha(\tau) + \alpha_0) & \ -\sin(-\alpha(\tau) + \alpha_0) & \ 0  & \ -t^\text{x}(\tau) \\
		\sin(-\alpha(\tau) + \alpha_0) &  \cos(-\alpha(\tau) + \alpha_0) & \ 0  & \ -t^\text{y}(\tau) \\
		-m^\text{x}(\tau)          &  -m^\text{y}(\tau)          & \ 1  & \ -t^\text{z}(\tau) 			\end{pmatrix}
	\begin{pmatrix}
		\stackrel[\displaystyle |]{\displaystyle |}{\vec x} \\
						1
	\end{pmatrix}.
\end{equation}
Here,
$\vec x$ and $\tau$ are the 3D position and acquisition time of a voxel and $\vec p = (\vec t^\text{x}, \vec t^\text{y}, \vec t^\text{z}, \vec m^\text{x}, \vec m^\text{y}, \vec \alpha)$ is the vector of motion parameters for all scans. $\vec t^\text{x/y/z}$ is the transverse (x,y) and axial (z) displacement of the retina.
Depending on the fast scan direction, either $\vec m^\text{x}$ or $\vec m^\text{y}$ is nonzero and describes axial shearing of B-scans originating from scanning beam to pupil center alignment~\cite{Kraus2014}. The transverse rotation $\vec \alpha$ corresponds to torsional motion~\cite{Lezama2016}, which we assume constant within each scan. For all other parameter types, parameters (e.\,g.\ $\vec t^\text{x}$) are estimated
for the time points corresponding to the centers of each B-scan
repeat
and then interpolated
along time using a cubic hermite spline
($t^\text{x}(\tau)$) to attain A-scan-specific values.
To reduce displacement bias from equidistant resampling of the registration target grid to the axes-aligned moving B-scans~\cite{Aganj2013}, a constant $\alpha_0 \approx 45^\circ$ is added.
Lastly, besides other effects, illumination can vary with beam alignment / motion,
and potentially confound registration.
Therefore, based on a (log-scale) voxel intensity $s$, change in illumination is modeled as
\begin{equation}
	\mathcal I(s,\tau,\vec c) = s + I(s > s_\text{min}) \cdot c(\tau),
\end{equation}
where $c(\tau)$ is a spline given by parameters $\vec c$, $I$ is the indicator function, and $s_\text{min}$ is a threshold to ignore the background, which is not affected by illumination.

Parameters are optimized jointly by minimization of the objective function
\begin{equation}
	\mathcal J(\vec p, \vec c) = \sum\nolimits_{M \in \mathcal V} \mathcal D(\vec p, \vec c, M) + \mathcal R(\vec p, \vec c) \quad \text{s.t.} \quad \mathcal C(\vec p, \vec c) = \vec 0,
\end{equation}
comprised of data terms $\mathcal D$ for all volumes $M \in \mathcal V$, and a smoothness regularization term $\mathcal R$, which penalizes the squared difference between sequential parameters with parameter type-specific weighting factors.
Optimization is performed via momentum gradient descent with parameter type-specific step sizes until the maximum change
falls below a threshold.
The constraint $\mathcal C$ enforces parameters to have zero mean (by mean subtraction after each optimizer step).
Optimization is performed in a 4-level coarse-to-fine multi-resolution pyramid of the (preprocessed) input, created by consecutive factor 2 downsamplings in axial direction.
Axial displacement parameters are initialized to the average depth of the voxels of temporally neighboring A-scans, weighted by their cubed intensity. This
aligns the bright horizontal band corresponding to the retinal pigment epithelium (see B-scan in \figurename~\ref{fig:oct-scanning}).
Other parameters are initialized with zeros.

Each data term $\mathcal D$ operates on the illumination-corrected voxels $\tilde s_{i,k}^M(\vec c) := \mathcal I(s_{i,k}^M,\tau_i^M,\vec c)$ of a moving volume $M$,
where $i$, $k$, $s_{i,k}^M$ and $\tau_i^M$ are a voxel's A-scan index, depth index, intensity and acquisition time.
To compute the corresponding squared difference loss for each target $T$, target volumes $\tilde S^T$ are interpolated to the voxel's motion-corrected location $\vec{\tilde x}_{i,k}^M(\vec p) := \mathcal T(\vec x_{i,k}^M, \tau_i^M, \vec p)$,
where $\vec x_{i,k}^M$ is the original voxel position,
via 3D cubic hermite spline interpolation $\mathcal W$:
\begin{equation}
	\mathcal D(\vec p, \vec c, M) = \sum_{\substack{T \in \mathcal V \\ \dir(T) \neq \dir(M)}} \sum_{i = 1}^{wh} \sum_{\substack{k = 1 \\ \valid\left(\tilde S^T, \vec{\tilde x}_{i,k}^M(\vec p)\right)}}^d \left( \tilde s_{i,k}^M(\vec c) - \mathcal W\big(\tilde S^T, \vec{\tilde x}_{i,k}^M(\vec p) \big) \right)^2
\end{equation}
Only volumes with orthogonal B-scan orientation (given by $\dir(\cdot)$) are used as targets.
$wh$ and $d$ are the number of A-scans and their depth.
Target volumes $\tilde S^T$ are assumed constant within each data term evaluation, to limit computational demands of the gradient evaluation.
The implementation for forward-warping the targets' A-scans by $\mathcal T$ is detailed later.
Due to use of forward warping, gaps in $\tilde S^T$ are known. Consequently, $\mathcal W$ is only defined if the $4^3$ neighborhood around the moving voxel's location is valid, and ignored otherwise, as determined by $\valid(\tilde S^T, \vec x)$.
In the final multi-resolution level, the target volume is computed with similar resolution as the preprocessed input, and is reduced in factor 2 steps in all dimensions for the smaller levels.
Again, to avoid displacement bias during resampling~\cite{Aganj2013}, pseudo-random subpixel offsets are introduced by a slightly lower transverse resolution and z-position offsets (detailed in supplementary \figurename~\ref{fig:aliasing}).

Besides accuracy, a primary concern for algorithm design was a small memory footprint, to allow joint registration of many volumes on a single GPU. Under these preconditions,
the algorithm was designed for massively parallel computing throughout and implemented in CUDA 11.1.
The
most
demanding
step is the forward-resampling of the target volume.
Naive mapping to a $4^3$ neighborhood in the target grid necessitates complex computations to determine each voxel-specific weight, rendering the method infeasible for clinical use. We utilize the absence of axial distortion within A-scans in two critical ways: First, we perform a cubic hermite spline-weighted warping only in axial direction of each A-scan, such that data points are axially aligned with the target grid voxels. This limits the complexity of scattered-data warping to the 2 transverse dimensions only, where we use a truncated Gaussian weighting (sidelength 4, $\sigma = 0.5$, in target
pixels). Secondly, we precompute and share coefficients across the axial direction. For a volume with sidelength $N$, separability and precomputation dramatically reduce the number of coefficient computations from $N^3 \cdot 4^3$ to $N^2 \cdot (4 + 4^2)$.

\section{Experiments and Results} \label{results}
A dataset of 18 patients was acquired on a prototype spectral domain OCT scanner with 128 kHz A-scan rate over a $6 \times 6$ mm field with $500 \times 500$ A-scans without B-scan repeats. The axial resolution was 3~\textmu m FWHM, axial pixel spacing was 1.78~\textmu m, A-scans had 775 pixels. For each subject, in one eye, 4 volumes were acquired at the fovea with alternating B-scan orientation.
Eye tracking was not available.
The study cohort is detailed in the supplementary material, \tablename~\ref{tab:data}.
We used
6 subjects to tune hyperparameters, mainly step size and convergence tolerance, and report our results on the remaining 12 subjects.

Evaluation of OCT motion correction accuracy is challenging~\cite{Brea2019}, ground truth motion is not available. Discontinuities limit the definition of landmarks, as well as error computation between backward displacements, to pixel accuracy. Positional ambiguities arise due to overlaps/gaps. This doesn't apply to the axial direction, which was evaluated in~\cite{Kraus2014} via reproducibility of segmented retinal layers, but localization accuracy
($\sim$3~\textmu m~\cite{He2019}) is worse than registration accuracy. 
In contrast, forward-warped displacements allow \emph{direct} computation of exact differences.
We register each scan to both orthogonally acquired scans \emph{independently}.
Local distortion is independent from the coregistered scan used, so both estimated displacement fields should be identical up to global differences from scanner-to-eye alignment,
which must be compensated. We used two transforms: Eq.~\eqref{eq:transform}, fixed to a single time-point $\tau$, only removes \emph{rigid} movement in the optical setup, and describes (all) residual distortion. The \emph{affine-like} transform removes two further types of global linear distortion, leaving only local distortion and providing a descriptor of misalignment.
This transform, and prevention of bias from resampling effects, are described in the \hyperref[sec:supplementary]{supplementary material}.

\begin{figure}[p]
	\centering
	\includegraphics{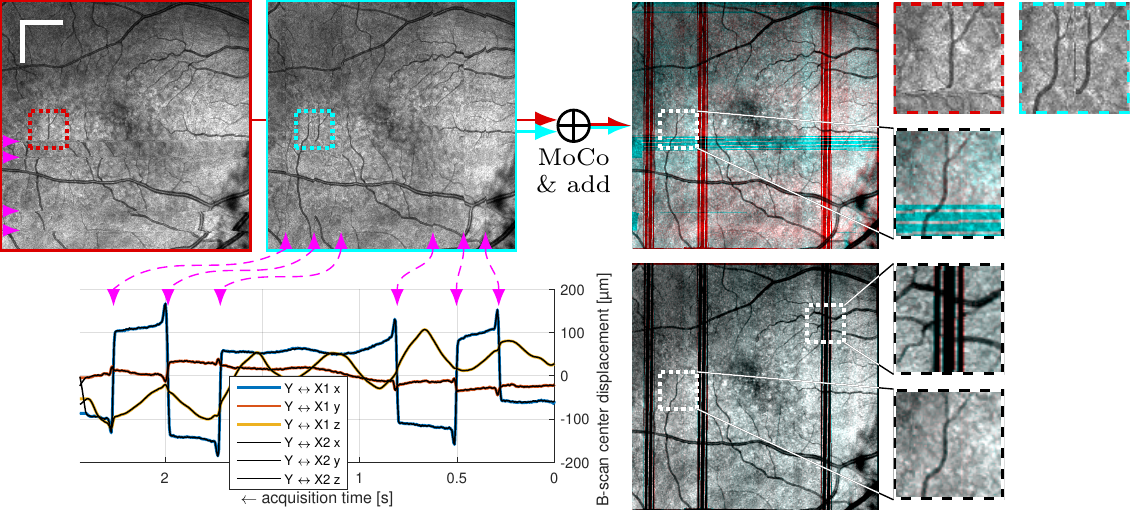}
	\caption{Top: Input X- and Y-fast en face images, and red/cyan composite image after motion correction (MoCo). Bottom: Red/cyan composite of a scan Y registered to tar\-gets X1/X2, after affine-like alignment.
	Plot: The $x,y,z$ displacements of the B-scan centers, computed with X1 (colored) and X2 (black),
		as compared in the quantitative evaluation (\figurename~\ref{fig:quantitative}).
	Saccades consistently overshoot in this subject.
	Scalebars 1\,mm.}							\label{fig:qualitative}\end{figure}\begin{figure}[p]
	\centering
	\includegraphics[keepaspectratio,height=3cm]{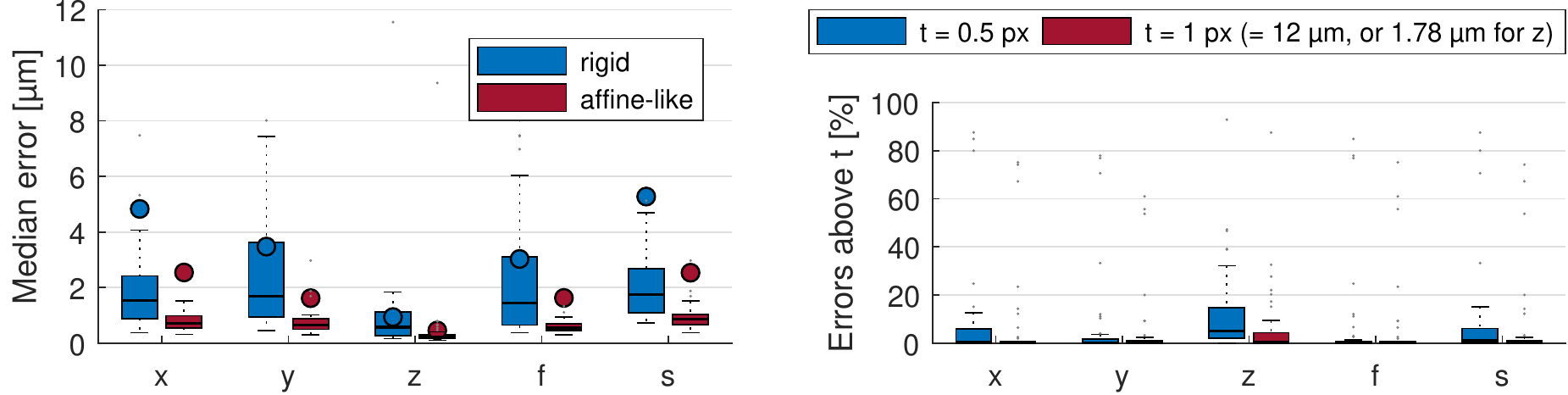}
	\caption{Box plot and means (circles) of scan reproducibility metrics in x, y, z, fast and slow direction. Left: Median A-scan displacement distance. The y-axis range
		is
	one transverse pixel spacing. Right: Fraction of distances $> 0.5$ and 1 px (affine-like).}
	\label{fig:quantitative}
\end{figure}
\begin{figure}[p]
\centering
\includegraphics{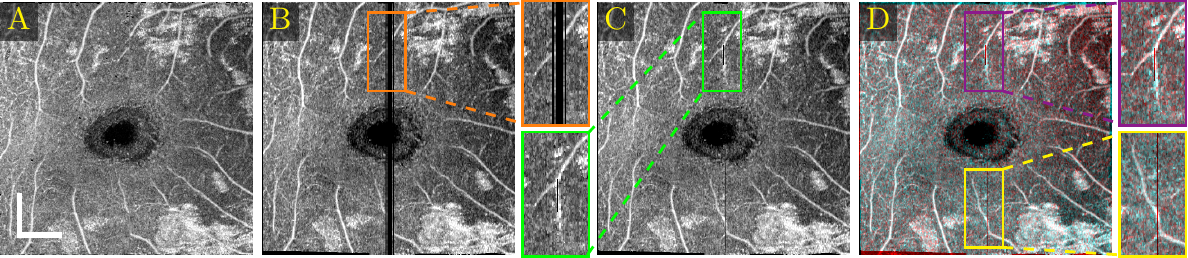}
\caption{Demonstration of individual B-scan repeat registration and gap filling in an example swept-source OCT dataset with 8 B-scan repeats. A: Uncorrected Y-fast volume, projected over the superficial capillary plexus. White spots originate from nerve fiber layer segmentation inaccuracies. B: Corrected using only 1 (constant) displacement parameter set for all B-scan repeats, limiting repeated A-scans to a single position, like previous methods. A gap appears at a slow scan direction microsaccade. C: Registration with individual displacements for each A-scan-repeat reveals their true arrangement: The repeated acquisitions cover the gap almost uniformly. D: The red/cyan composite X/Y-fast image demonstrates correctness of the arrangement in C. Scalebars 1\,mm.
}
\label{fig:rep-bscans}
\end{figure}
Images and estimated displacements are shown for a representative scan in \figurename~\ref{fig:qualitative}.
Tiny discontinuities prove absence of overregularization, and consistent transverse vibration indicates partial correction of ocular tremor.
For quantitative analysis, we computed
the median distance between the aligned A-scan displacements, and the fraction of displacements with a distance above 0.5 (problematic for supersampling) and 1 pixels (misalignments).
The first and last 5\% of B-scans were excluded, because they might not overlap with the orthogonal data, preventing
registration.
As the distributions are heavily skewed, we present box plots in \figurename~\ref{fig:quantitative}. The three outliers in each direction in the right plot originate from the same subject, which is shown in supplementary \figurename~\ref{fig:aliasing}.
It is critical to note that the parameter density (B-scan rate 205 Hz) of the hermite splines is insufficient to fully correct ocular tremor (frequency up to $\sim$100 Hz~\cite{MartinezConde2004}). 
Therefore, this
aperiodic,
wave-like motion (amplitude $\sim$30$''$ $\equiv$ $\sim$1.6~\textmu m on the retina~\cite{MartinezConde2004}) cannot be fully corrected, and neither is fully represented in the reproducibility error. In the transverse directions, this puts a lower accuracy limit on the evaluation scheme, but it is small compared to the pixel spacing (12~\textmu m).
Using an Nvidia RTX 5000 GPU, the median and maximum runtime in the test set, excluding disk I/O which is irrelevant in clinical routine, was 8.6~s and 31.3~s. \tablename~\ref{tab:runtime} compares average run\-times of various methods.
Lastly, registration of individual B-scan repeats is demonstrated on a swept-source scanner in \figurename~\ref{fig:rep-bscans}.

\begin{table}[t]
\caption{Average runtime
for registering $\sim$$2 \times 500^2$ A-scans. Not all prior work reported runtime.
$^1$I/O
time not reported, we subtracted 10 s. $^2$2D transverse correction only.}\label{tab:runtime}
\centering
\begin{tabular}{|l|c|c|c|c|c|}\hline
 & Zang ('17)\,\cite{Zang2017}$^1$ & Ploner ('21)\,\cite{Ploner2021} & Makita ('21)\,\cite{Makita2021}$^{1,2}$ & Cheng ('21)\,\cite{Cheng2021}$^1$ & ours \\ \hline
Runtime & 6.83 min & 84 s & $\ge 2.32$ min & 5.83 min & \textbf{15 s} \\ \hline
\end{tabular}
\end{table}

\section{Discussion}
Using our spatiotemporal model, we achieve robust estimation of 3D motion traces of the retina in pathology even during fast microsaccadic eye movements.
Registration accuracy is improved dramatically, to sub-micron residual axial distortion (0.51~\textmu m median of medians) and even smaller local distortion (misalignment) of 0.20~\textmu m median of medians.
Transverse errors are only limited by ocular tremor (peak-to-peak amplitude less than $3.2$~\textmu m $\approx \frac{1}{4}$ px).
At the same time, on single- or merged-B-scan-repeat data, the runtime of our approach outperforms published methods by a factor of 5 and more.
Whereas runtime increases with additional B-scan repeats, the necessity for prior registration of repeated B-scans is removed.
We presented the first
metric/distance-based 3D error quantification in OCT motion correction. Due to lack of such evaluation of previous methods, a direct comparison is not possible (see section~\ref{results}). However, by investigating algorithm properties, lower bounds for accuracy can be derived: Athwal et al.\ create a registration target by stitching saccade-free segments~\cite{Athwal2021}. Drift motion in the reference segment is not corrected, resulting in residual transverse distortion in the range of 6$'$ $\approx$ $\sim$19.1~\textmu m on the retina~\cite{MartinezConde2004} and alignment is computed in whole-pixel steps only (12~\textmu m in our dataset, typically $\ge 6$~\textmu m).
Axial registration is limited by segmentation accuracy (ca.\ 3~\textmu m~\cite{He2019}) and reliability is limited in pathology~\cite{Athwal2021}.
In~\cite{Kraus2014} and~\cite{Ploner2021}, performing the B-scan shear estimation before transverse motion compensation puts a limit on overall axial registration accuracy. 7.0 and 3.7~\textmu m were reported for axial reproducibility and misalignment errors. However, the evaluation compared retinal layer positions, whose segmentation introduces additional error.
As detailed in the introduction, these methods are further limited by model inaccuracies
in the transverse directions, and are unreliable near microsaccades~\cite{Ploner2021}. Incompatibility with head rotation limits applicability to many-volume and widefield registration~\cite{Lezama2016}. Displacement bias from resampling was not discussed in
prior 3D OCT registration literature.

The advances of our model open up various new possibilities for exploration that go beyond motion correction: Subpixel-accuracy and direct forward warping allow \emph{supersampling} using repeated volume scans~\cite{Greenspan2009}, enabling a new paradigm for high-density widefield OCT imaging that is more robust to saccadic eye motion.
OCT-derived blood flow signals like OCTA~\cite{Spaide2018} can not only be merged and analyzed at capillary vasculature scale, but the maintained temporal dependency allows cardiac cycle-aware \emph{4D spatiotemporal analysis and reconstruction} of flow speeds via Doppler-OCT~\cite{Leitgeb2014} and VISTA-OCTA~\cite{Ploner2016}.
For both OCT intensity and derived signals, \emph{advanced 3D signal reconstruction} allows probabilistic maximum a-posteriori modeling of the non-Gaussian noise distribution~\cite{DuBose2018} in the reconstruction loss, or deblurring~\cite{Farsiu2004}, \textit{directly based on raw voxel data}.
The reconstructed eye motion trace extends OCT with potential for
accessible \emph{screening of neurological diseases} that manifest in oculomotor dysfunction~\cite{Fletcher1986,Gitchel2012,Mallery2018}. 

Given the variety of ocular pathologies and OCT scanners, performance should be reaffirmed in a larger cohort.
Hardware tracking could be included to improve reliability and prevent longer runtimes in subjects with severe motion.
To make optimal use of the motion correction, we are working on
an
iterative reconstruction and plan
a release of the complete framework in the future.
\subsubsection{Acknowledgements.} {DFG MA 4898/12-1, \color{black} NIH 5-R01-EY011289-35.}

\vfill \pagebreak 
\bibliographystyle{splncs04}
\bibliography{literature}

\begin{thebibliography}{10}
\providecommand{\url}[1]{\texttt{#1}}
\providecommand{\urlprefix}{URL }
\providecommand{\doi}[1]{https://doi.org/#1}

\bibitem{Aganj2013}
Aganj, I., Yeo, B.T.T., Sabuncu, M.R., Fischl, B.: On removing interpolation
  and resampling artifacts in rigid image registration. IEEE Transactions on
  Image Processing  \textbf{22}(2),  816--827 (2013).
  \doi{10.1109/TIP.2012.2224356}

\bibitem{Athwal2021}
Athwal, A., Balaratnasingam, C., Yu, D.Y., Heisler, M., Sarunic, M., Ju, M.:
  Optimizing {3D} retinal vasculature imaging in diabetic retinopathy using
  registration and averaging of {OCT-A}. Biomed. Opt. Express  \textbf{12}(1),
  553--570 (2021). \doi{10.1364/BOE.408590}

\bibitem{Cheng2021}
Cheng, Y., Chu, Z., Wang, R.K.: Robust three-dimensional registration on
  optical coherence tomography angiography for speckle reduction and
  visualization. Quantitative Imaging in Medicine and Surgery  \textbf{11}(3)
  (2021). \doi{10.21037/qims-20-751}

\bibitem{DuBose2018}
Dubose, T.B., Cunefare, D., Cole, E., Milanfar, P., Izatt, J.A., Farsiu, S.:
  Statistical models of signal and noise and fundamental limits of segmentation
  accuracy in retinal optical coherence tomography. IEEE Transactions on
  Medical Imaging  \textbf{37}(9),  1978--1988 (2018).
  \doi{10.1109/TMI.2017.2772963}

\bibitem{Farsiu2004}
Farsiu, S., Robinson, M., Elad, M., Milanfar, P.: Fast and robust multiframe
  super resolution. IEEE Transactions on Image Processing  \textbf{13}(10),
  1327--1344 (2004). \doi{10.1109/TIP.2004.834669}

\bibitem{Fletcher1986}
Fletcher, W.A., Sharpe, J.A.: Saccadic eye movement dysfunction in alzheimer's
  disease. Annals of Neurology  \textbf{20}(4),  464--471 (1986).
  \doi{10.1002/ana.410200405}

\bibitem{Gitchel2012}
Gitchel, G.T., Wetzel, P.A., Baron, M.S.: {Pervasive Ocular Tremor in Patients
  With Parkinson Disease}. Archives of Neurology  \textbf{69}(8),  1011--1017
  (2012). \doi{10.1001/archneurol.2012.70}

\bibitem{Greenspan2009}
Greenspan, H.: {Super-Resolution in Medical Imaging}. The Computer Journal
  \textbf{52}(1),  43--63 (2008). \doi{10.1093/comjnl/bxm075}

\bibitem{He2019}
He, Y., Carass, A., Liu, Y., Jedynak, B.M., Solomon, S.D., Saidha, S.,
  Calabresi, P.A., Prince, J.L.: Fully convolutional boundary regression for
  retina oct segmentation. In: Shen, D., Liu, T., Peters, T.M., Staib, L.H.,
  Essert, C., Zhou, S., Yap, P.T., Khan, A. (eds.) Medical Image Computing and
  Computer Assisted Intervention -- MICCAI 2019. pp. 120--128. Springer
  International Publishing (2019). \doi{10.1007/978-3-030-32239-7_14}

\bibitem{Huang1991}
Huang, D., Swanson, E., Lin, C., Schuman, J., Stinson, W., Chang, W., Hee, M.,
  Flotte, T., Gregory, K., Puliafito, C., Fujimoto, J.: Optical coherence
  tomography. Science  \textbf{254}(5035),  1178--1181 (1991).
  \doi{10.1126/science.1957169}

\bibitem{Kraus2014}
Kraus, M., Liu, J.J., Schottenhamml, J., Chen, C.L., Budai, A., Branchini, L.,
  Ko, T., Ishikawa, H., Wollstein, G., Schuman, J., Duker, J., Fujimoto, J.,
  Hornegger, J.: Quantitative {3D-OCT} motion correction with tilt and
  illumination correction, robust similarity measure and regularization.
  Biomed. Opt. Express  \textbf{5}(8),  2591--2613 (2014).
  \doi{10.1364/BOE.5.002591}

\bibitem{Leitgeb2014}
Leitgeb, R.A., Werkmeister, R.M., Blatter, C., Schmetterer, L.: Doppler optical
  coherence tomography. Progress in Retinal and Eye Research  \textbf{41},
  26--43 (2014). \doi{10.1016/j.preteyeres.2014.03.004}

\bibitem{Lezama2016}
Lezama, J., Mukherjee, D., McNabb, R., Sapiro, G., Kuo, A., Farsiu, S.:
  Segmentation guided registration of wide field-of-view retinal optical
  coherence tomography volumes. Biomed. Opt. Express  \textbf{7}(12),
  4827--4846 (2016). \doi{10.1364/BOE.7.004827}

\bibitem{Makita2021}
Makita, S., Miura, M., Azuma, S., Mino, T., Yamaguchi, T., Yasuno, Y.:
  Accurately motion-corrected lissajous {OCT} with multi-type image
  registration. Biomed. Opt. Express  \textbf{12}(1),  637--653 (2021).
  \doi{10.1364/BOE.409004}

\bibitem{Mallery2018}
Mallery, R.M., Poolman, P., Thurtell, M.J., Full, J.M., Ledolter, J.,
  Kimbrough, D., Frohman, E.M., Frohman, T.C., Kardon, R.H.: {Visual Fixation
  Instability in Multiple Sclerosis Measured Using SLO-OCT}. Investigative
  Ophthalmology \& Visual Science  \textbf{59}(1),  196--201 (2018).
  \doi{10.1167/iovs.17-22391}

\bibitem{MartinezConde2004}
Martinez-Conde, S., Macknik, S., Hubel, D.: The role of fixational eye
  movements in visual perception. Nat Rev Neurosci  \textbf{5}(3),  229--240
  (2004). \doi{10.1038/nrn1348}

\bibitem{Ploner2021}
Ploner, S.B., Kraus, M., Moult, E., Husvogt, L., Schottenhamml, J., Alibhai,
  A., Waheed, N., Duker, J., Fujimoto, J., Maier, A.K.: Efficient and high
  accuracy {3-D} {OCT} angiography motion correction in pathology. Biomed. Opt.
  Express  \textbf{12}(1),  125--146 (2021). \doi{10.1364/BOE.411117}

\bibitem{Ploner2016}
Ploner, S.B., Moult, E.M., Choi, W., Waheed, N.K., Lee, B., Novais, E.A., Cole,
  E.D., Potsaid, B., Husvogt, L., Schottenhamml, J., Maier, A., Rosenfeld,
  P.J., Duker, J.S., Hornegger, J., Fujimoto, J.G.: Toward quantitative optical
  coherence tomography angiography. Retina  \textbf{36},  S118--S126 (2016).
  \doi{10.1097/IAE.0000000000001328}

\bibitem{Schwarzhans2020}
Schwarzhans, F., Desissaire, S., Steiner, S., Pircher, M., Hitzenberger, C.K.,
  Resch, H., Vass, C., Fischer, G.: Generating large field of view en-face
  projection images from intra-acquisition motion compensated volumetric {OCT}
  data. Biomed. Opt. Express  \textbf{11}(12),  6881--6904 (2020).
  \doi{10.1364/BOE.404738}

\bibitem{Spaide2018}
Spaide, R., Fujimoto, J., Waheed, N., Sadda, S., Staurenghi, G.: Optical
  coherence tomography angiography. Progress in Retinal and Eye Research
  \textbf{64},  1--55 (2018). \doi{10.1016/j.preteyeres.2017.11.003}

\bibitem{Brea2019}
Sánchez~Brea, L., Andrade De~Jesus, D., Shirazi, M.F., Pircher, M., van
  Walsum, T., Klein, S.: Review on retrospective procedures to correct retinal
  motion artefacts in {OCT} imaging. Applied Sciences  \textbf{9}(13) (2019).
  \doi{10.3390/app9132700}

\bibitem{Zang2017}
Zang, P., Liu, G., Zhang, M., Wang, J., Hwang, T.S., Wilson, D.J., Huang, D.,
  Li, D., Jia, Y.: Automated three-dimensional registration and volume
  rebuilding for wide-field angiographic and structural optical coherence
  tomography. {J. Biomed. Opt.}  \textbf{22}(2),  26001 (2017).
  \doi{10.1117/1.JBO.22.2.026001}

\end{thebibliography}
\vfill\pagebreak
\section*{Supplementary Materials}\label{sec:supplementary}
\subsubsection{Global alignment.}
%
\begin{equation}
	\mathcal T_\text{affine-like}(x, y, z) = \begin{pmatrix}
		a_\text{xx} & a_\text{xy} & 0 & 0           & t_\text{x} \\
		a_\text{yx} & a_\text{yy} & 0 & 0           & t_\text{y} \\
		m_\text{x}  & m_\text{y}  & 1 & m_\text{xy} & t_\text{z}
	\end{pmatrix}
	\begin{pmatrix}
		x \\
		y \\
		z \\
		xy \\
		1
	\end{pmatrix} \label{eq:affine-like}
\end{equation}
Transverse ($a_{...}$) and diagonal axial sheering ($m_\text{xy}$) are challenging to correct using orthogonal raster-scans, and constitute most residual distortion. This distortion
is consistent among
jointly registered scans.
When comparing independently corrected scans, e.\,g.\ for follow-up,
$\mathcal T_\text{affine-like}$
can be used for global alignment, making this distortion clinically irrelevant.
After compensation, only local distortions remain. Due to orthogonal acquisition, these must be misalignments.

\subsubsection{Potential bias during evaluation.}
\begin{figure}[t]
\includegraphics[keepaspectratio,width=.29\textwidth]{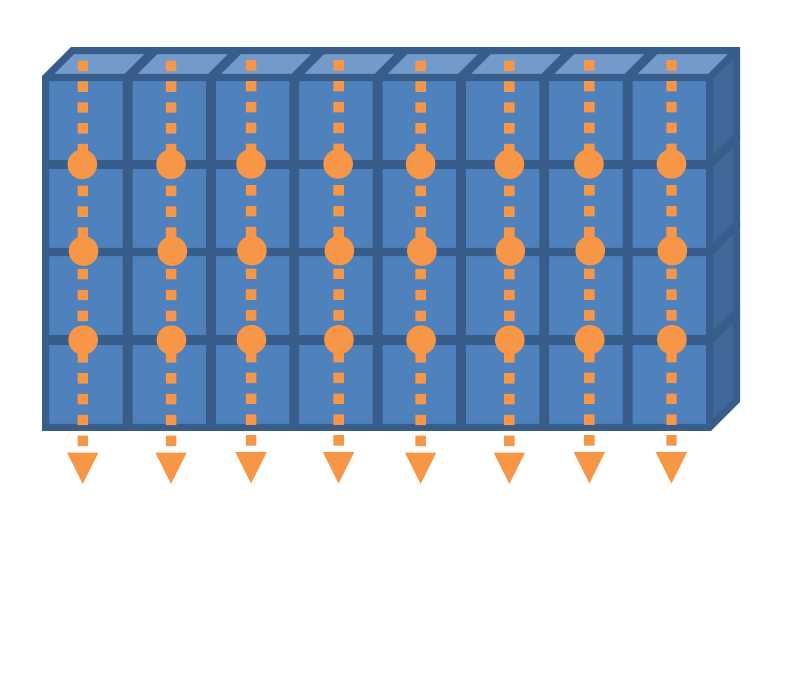} \hfill
\includegraphics[keepaspectratio,width=.29\textwidth]{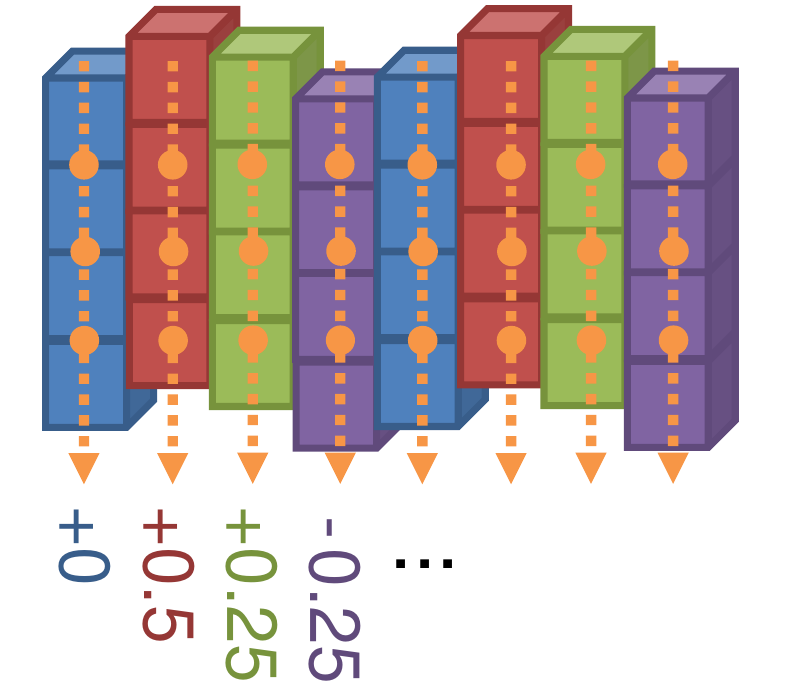} \hfill
\includegraphics[keepaspectratio,width=.29\textwidth]{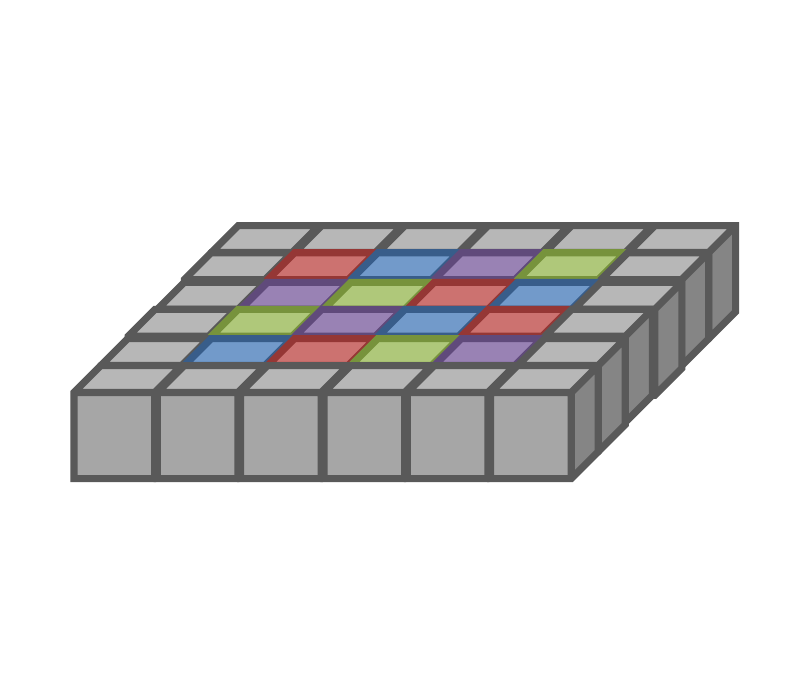}
\caption{Compensation of bias from uniform axial resampling of the target grid to the moving B-scan voxel positions~\cite{Aganj2013}. Left: Naive approach. Sampling points (orange dots) have equal offsets from the target grid (cubes). Biases add up throughout the B-scan, causing artificial local minima that can prevent subpixel registration. Center: Target grid axial positions have pseudo-random offsets. Bias effects cancel each other out. Right: Offset permutations in an en face plane. The $4 \times 4$ pattern is repeated throughout.
}
\label{fig:aliasing}
\end{figure}

Registration is susceptible to overly favor solutions where the data is blurred by the interpolator, as can happen when sampling between data points, as this reduces the inconsistency among noisy samples and, consequently, the objective function. Aliasing effects during equidistant resampling of moving, interpolated targets add to this effect~\cite{Aganj2013} (see also supplementary \figurename~\ref{fig:aliasing}).
Although we did not observe this in the estimated motion traces, such effects could cause displacements to change not continuously, but ``quantized'' in pixel-steps.
For identical registration targets, such artifactual steps could align consistently and therefore result in small reproducibility errors despite up to half-pixel residual distortion.
Using the pseudo-random resampling that prevents these effects between the moving and target volumes in our method, we prohibit these effects also in the reproducibility evaluation, by using different target grid configurations in the two motion-corrections used to generate the compared displacement fields. We used $\alpha_0 = \frac\pi5$ / $\frac\pi6$ ($\not\approx 45^\circ$), resolution factor $\frac56$ / $\frac45$ ($\lesssim$ 1), and axial offsets (supplementary \figurename~\ref{fig:aliasing}) were offset by $(0,0)$ / $(2,2)$ in transverse directions. Potential pixel-steps in the displacements would not be aligned, and, after averaging, result in the expected error.
\subsubsection{Unit conversions.} For conversion between rotation angles and distance on retinal surface, we assume a spherical retina with 22\,mm diameter.
\subsubsection{Author Contributions.} {SP is the main author. Algorithm design, evaluation, implementations, initial writing: SP. Scanner design: SC. Data: JW, SC. Result analysis and presentation: SP, KB, SC, JW. Funding: AM, SP, JF. All authors
contributed to the final manuscript.}

\begin{table}
\caption{Study cohort. Data collection was performed at the New England Eye Center (NEEC), Boston, USA, in agreement with the declaration of Helsinki and in accordance with the institutional review boards at Massachusetts Institute of Technology and NEEC.
The only exclusion criteria were severe defocus / illumination artifacts that would prevent clinical use, or the retina being outside the imaging range for significant portions of the scan. Assignment to train / test set was performed based on metadata to balance age and gender distributions, by a person who had not seen the images.}
\label{tab:data}
\centering
\begin{tabular}{|l|l|c|c|c|c|}
\hline
Set & Pathology & \#patients & \#eyes & Age & Female \\\hline
Train & Age-related macular degeneration & 3 & 3 & $76.66 \pm \phantom{0}5.73$ & 66\% \\
      & Non-proliferative diabetic retinopathy & 3 & 3 & $65\phantom{.00} \pm 11.05$ & 33\% \\\hline
Test  & Age-related macular degeneration & 6 & 6 & $74.16 \pm \phantom{0}8.90$ & 50\% \\
      & Non-/proliferative diabetic retinopathy & 6 & 6 & $63.83 \pm 10.48$ & 50\% \\
\hline
\end{tabular}
\end{table}
\begin{figure}[t]
\centering
\begin{tikzpicture}[inner sep=0,outer sep=0,node distance=.1cm]
\tikzset{
	letter/.style 2 args = {
		append after command={(\tikzlastnode.north west) node[anchor=north west,text=#2,inner sep=0,inner xsep=2](){\strut #1}} 
	},
	filledletter/.style n args={4}{
		append after command={(\tikzlastnode.north west) node[anchor=north west,text=#2,fill=#3,fill opacity=#4,text opacity=1,inner sep=0,inner xsep=2](){\strut #1}}
	},
	letter/.default={a}{black},
	filledletter/.default={a}{black}{white}{1}
}
\let\picwidth\relax
\newlength\picwidth
\setlength{\picwidth}{.2\textwidth}
\node[filledletter= {X1}{yellow}{black}{0.5}             ] (X1){\includegraphics[keepaspectratio,width=\picwidth]{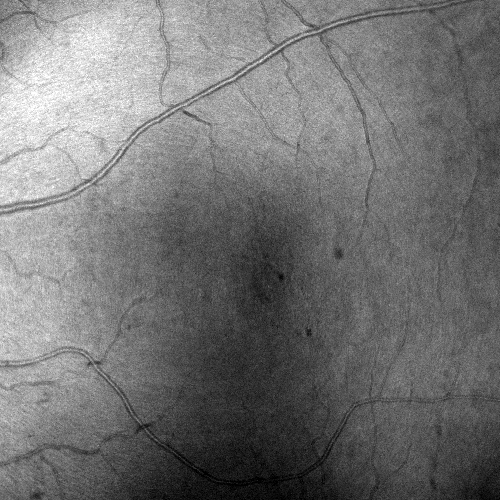}};
\node[filledletter= {Y1}{yellow}{black}{0.5},right=of  X1] (Y1){\includegraphics[keepaspectratio,width=\picwidth]{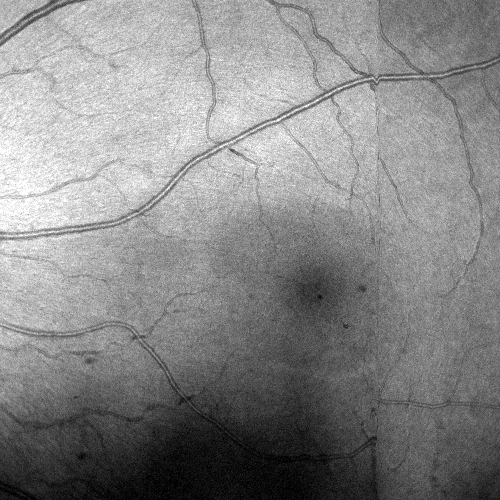}};
\node[filledletter= {X2}{yellow}{black}{0.5},right=of  Y1] (X2){\includegraphics[keepaspectratio,width=\picwidth]{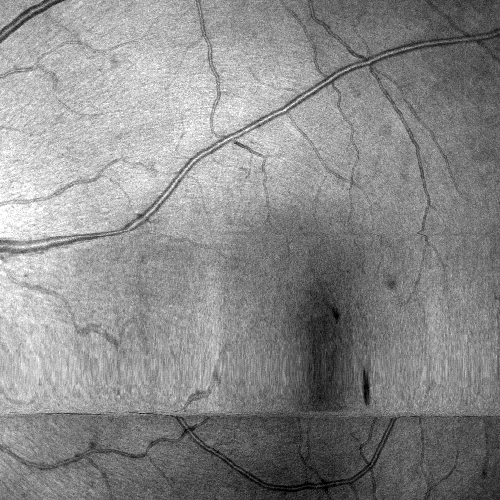}};
\node[filledletter= {Y2}{yellow}{black}{0.5},right=of  X2] (Y2){\includegraphics[keepaspectratio,width=\picwidth]{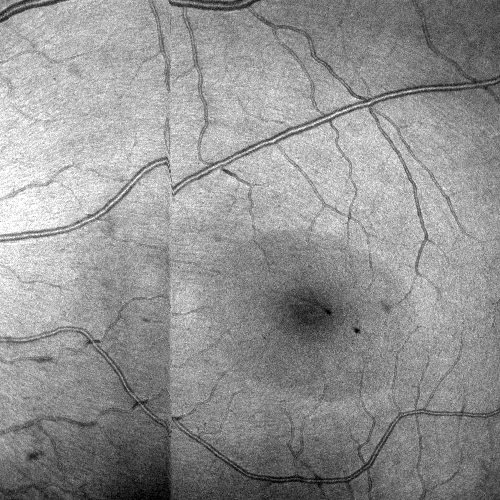}};
\setlength{\picwidth}{.2\textwidth}
\node[filledletter={X1 $\leftrightarrow$ Y2}{yellow}{black}{0.5},below right=.1cm and .35cm of Y1.south,anchor=north](C12){\includegraphics[keepaspectratio,width=\picwidth]{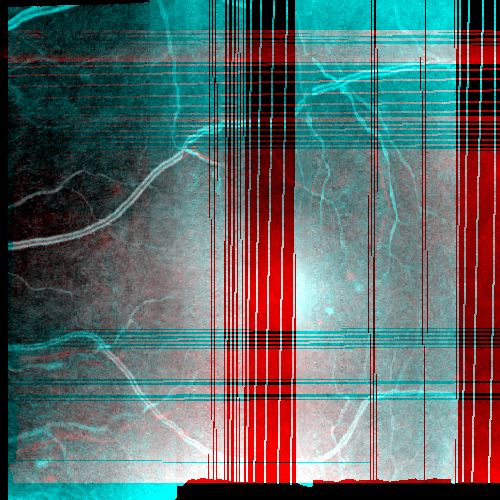}};
\draw[->,>=Latex,thick,purepink] ($(C12.south west)+(.23\picwidth,.08\picwidth)$) -- ++(-.04,.08cm);
\draw[->,>=Latex,thick,purepink] ($(C12.south west)+(.13\picwidth,.22\picwidth)$) -- ++(0.0,.1cm);
\draw[->,>=Latex,thick,orange] ($(C12.south west)+(.32\picwidth,.73\picwidth)$) -- ++(.1,0cm);
\draw[->,>=Latex,thick,purepink] ($(C12.south west)+(.58\picwidth,.66\picwidth)$) -- ++(.09,-.02cm);
%
\node[left=of C12.north west,anchor=north east,draw,purepink,ultra thick](C11Z){\includegraphics[keepaspectratio,height=.48\picwidth]{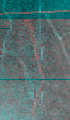}};
\draw[->,>=Latex,thick,purepink] ($(C11Z.south west)+(.2*.48\picwidth,.6*.48\picwidth)$) -- ++(.04,.08cm);
\node[left=of C12.south west,anchor=south east,draw,purepink,ultra thick](C12Z){\includegraphics[keepaspectratio,height=.48\picwidth]{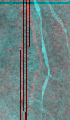}};
\draw[->,>=Latex,thick,purepink] ($(C12Z.south west)+(.2*.48\picwidth,.6*.48\picwidth)$) -- ++(.08,.04cm);
%
\node[filledletter={X1 $\leftrightarrow$ Y1}{yellow}{black}{0.5},left=of C11Z.north west,anchor=north east](C11){\includegraphics[keepaspectratio,width=\picwidth]{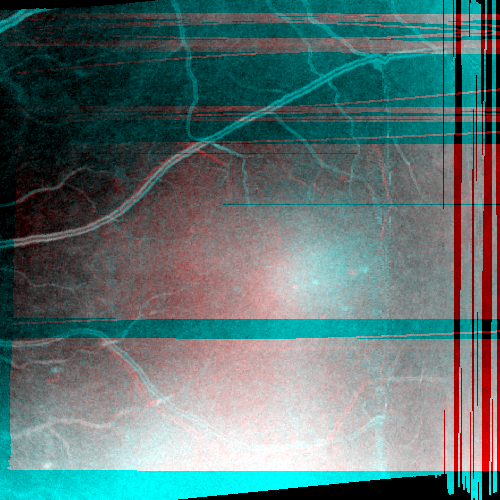}};
\draw[->,>=Latex,thick,orange] ($(C11.south west)+(.23\picwidth,.22\picwidth)$) -- ++(.04,.08cm);
\draw[->,>=Latex,thick,purepink] ($(C11.south west)+(.35\picwidth,.09\picwidth)$) -- ++(.04,.08cm);
\draw[->,>=Latex,thick,purepink] ($(C11.south west)+(.65\picwidth,.22\picwidth)$) -- ++(.08,-.04cm);
\draw[->,>=Latex,thick,orange] ($(C11.south west)+(.26\picwidth,.71\picwidth)$) -- ++(.072,-.072cm);
\draw[->,>=Latex,thick,purepink] ($(C11.south west)+(.40\picwidth,.63\picwidth)$) -- ++(.05,.09cm);
\draw[->,>=Latex,thick,purepink] ($(C11.south west)+(.57\picwidth,.59\picwidth)$) -- ++(.05,.09cm);
\draw[->,>=Latex,thick,orange] ($(C11.south west)+(.77\picwidth,.94\picwidth)$) -- ++(0.0,-.1cm);
\draw[thick, purepink] ($(C11.south west)+(.69\picwidth,.47\picwidth)$) rectangle ($(C11.south west)+(.84\picwidth,.76\picwidth)$);
\draw[dashed, purepink, thick] ($(C11.south west)+(.84\picwidth,.76\picwidth)$) -- (C11Z.north west);
\draw[dashed, purepink, thick] ($(C11.south west)+(.84\picwidth,.47\picwidth)$) -- (C11Z.south west);
\draw[thick, purepink] ($(C12.south west)+(.69\picwidth,.46\picwidth)$) rectangle ($(C12.south west)+(.84\picwidth,.72\picwidth)$);
\draw[dashed, purepink, thick] (C12Z.north east) -- ($(C12.south west)+(.69\picwidth,.72\picwidth)$);
\draw[dashed, purepink, thick] (C12Z.south east) -- ($(C12.south west)+(.69\picwidth,.46\picwidth)$);
\node[filledletter={X2 $\leftrightarrow$ Y1}{yellow}{black}{0.5},right=of C12](C21){\includegraphics[keepaspectratio,width=\picwidth]{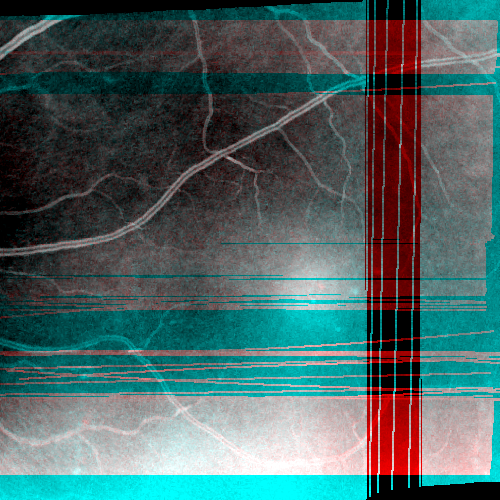}};
\node[filledletter={X2 $\leftrightarrow$ Y2}{yellow}{black}{0.5},right=of C21](C22){\includegraphics[keepaspectratio,width=\picwidth]{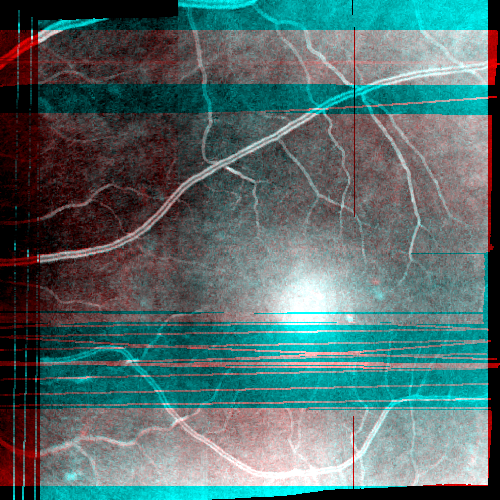}};
\end{tikzpicture}
\caption{The failed case. Top row: Input en face images.
Bottom row: Composite X-/ Y-fast motion corrected volumes in red/cyan, inverted to emphasize misregistrations. Both registrations with X1 show residual distortion (orange markers) and double vessels (pink markers). Notice that despite Y-fast volumes are well registered with X2, quantitative evaluation will report large discrepancy, because the displacements are inconsistent with the corresponding (bad) registrations with X1.}
\label{fig:outlier}
\end{figure}
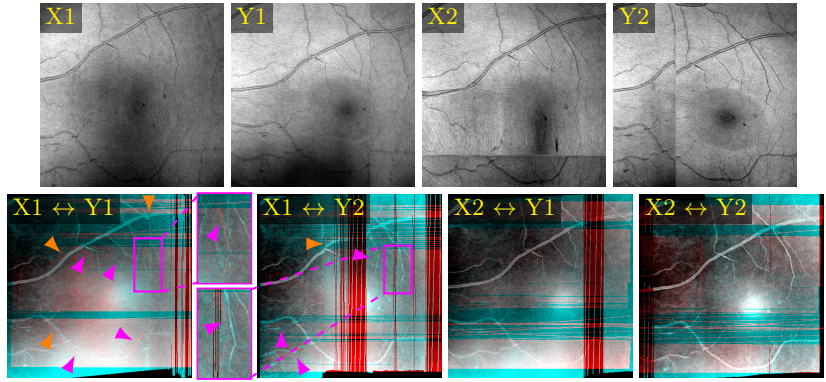
%
\end{document}